\documentclass[12pt]{article}
\usepackage{a4wide}
\usepackage{amssymb}
\begin{document} {\renewcommand{\thefootnote}{\fnsymbol{footnote}}

  \medskip \begin{center}
    {\LARGE The ground state of non-associative hydrogen\\ and upper bounds on
      the magnetic charge\\[2mm] of elementary particles}\\
    \vspace{1.5em}
Martin Bojowald,$^1$\footnote{e-mail address: {\tt bojowald@gravity.psu.edu}}
Suddhasattwa Brahma,$^2$\footnote{e-mail address: {\tt
    suddhasattwa.brahma@gmail.com}}
Umut B\"{u}y\"{u}k\c{c}am,$^1$\footnote{e-mail address: {\tt ubuyukcam@gmail.com}}\\
and Martijn van Kuppeveld$^1$\footnote{e-mail address: {\tt mvk18@psu.edu }}
\\
\vspace{0.5em}
$^1$ Institute for Gravitation and the Cosmos,
The Pennsylvania State
University,\\
104 Davey Lab, University Park, PA 16802, USA\\
\vspace{0.5em}
$^2$ Department of Physics, McGill University, Montr\'{e}al, QC H3A 2T8, Canada
\vspace{1.5em}
\end{center}
}

\setcounter{footnote}{0}

\begin{abstract}
  Formulations of magnetic monopoles in a Hilbert-space formulation of quantum
  mechanics require Dirac's quantization condition of magnetic charge, which
  implies a large value that can easily be ruled out for elementary particles
  by standard atomic spectroscopy. However, an algebraic formulation of
  non-associative quantum mechanics is mathematically consistent with
  fractional magnetic charges of small values. Here, spectral properties in
  non-associative quantum mechanics are derived, applied to the ground state
  of hydrogen with a magnetically charged nucleus. The resulting energy leads
  to new strong upper bounds for the magnetic charge of various elementary
  particles that can appear as the nucleus of hydrogen-like atoms, such as the
  muon or the antiproton.
\end{abstract}

\section{Introduction}

Eigenvalues and eigenstates can be defined and derived completely
algebraically, without using a Hilbert-space representation of observables as
operators. Such a formulation is important in particular in studies of
non-associative algebras that cannot be represented on a Hilbert
space. Physical examples can be found mainly in situations in which fractional
magnetic charges may be present that do not obey Dirac's quantization
condition \cite{DiracMonopoles}, which can be defined at the level of a
non-associative algebra of observables even though no Hilbert-space
representation exists \cite{Malcev,JackiwMon,Jackiw,OctQM}. Magnetic monopole
charges that obey Dirac's quantization condition are so large that they can
easily be ruled out in elementary particles by atomic spectroscopy.  While
small non-zero magnetic charges may be compatible with observational bounds,
they cannot obey the quantization condition and therefore require
non-associative algebras of observables.

Non-associative products are obtained for magnetic monopoles as follows: In
the presence of magnetic monopoles, the magnetic field has non-zero divergence
and therefore cannot be described by a vector potential. The usual canonical
momentum $\hat{\pi}_i=\hat{p}_i+e\hat{A}_i$ of a particle with electric charge
$e$ and mass $m$, where $\hat{p}_i=m\dot{\hat{x}}_i$ is the kinematical
momentum, is then unavailable. However, it turns out that the commutator of
two kinematical momenta,
\begin{equation} \label{pp}
 [\hat{p}_j,\hat{p}_k]=[\hat{\pi}_j-e\hat{A}_j, \hat{\pi}_k-e\hat{A}_k]=
i \hbar e \left(\widehat{\frac{\partial A_k}{\partial x_j}}-
  \widehat{\frac{\partial  A_j}{\partial x_k}}\right)=
i\hbar e\sum_{l=1}^3\epsilon_{jkl}
 \hat{B}^l
\end{equation}
does not require a vector potential. (The usual bracket
$[\hat{x}_j,\hat{p}_k]=i\hbar\delta_{jk}$ remains unchanged.) It can therefore
be generalized to a point charge moving in the presence of a background
magnetic charge, but it is not canonical and not even constant since the
magnetic field is position dependent. The Jacobi identity is therefore not
guaranteed to hold, and it is indeed violated as the calculation
\begin{equation} \label{ppp}
 [[\hat{p}_x,\hat{p}_y],\hat{p}_z]+  [[\hat{p}_y,\hat{p}_z],\hat{p}_x]+
 [[\hat{p}_z,\hat{p}_x],\hat{p}_y]
= i\hbar e \sum_{j=1}^3[\hat{B}^j,\hat{p}_j]= -\hbar^2e\;\widehat{{\rm
    div}\vec{B}}\not=0
\end{equation}
demonstrates. Since the assumption of an associative product would imply the
Jacobi idenity for the commutator, magnetic monopoles are seen to require
non-associative algebras of quantum observables
\cite{Malcev,JackiwMon,Jackiw,OctQM}. The basic commutators (\ref{pp})
together with an associator determined by (\ref{ppp}) can be turned into a
complete non-associative algebra by means of $*$-products
\cite{NonGeoNonAss,MSS1,BakasLuest,MSS2,MSS3}.

Purely algebraic derivations that do not make use of specific representations
are usually more challenging than standard quantum mchanics, in particular if
associativity cannot be assumed. As a consequence, such systems remain
incompletely understood, and it remains to be seen whether they can be
viable. Nevertheless, it has recently become possible to derive potential
physical effects \cite{NonAssEffPot} and to use spectral results for new upper
bounds on the possible magnetic charge of elementary particles
\cite{WeakMono}. The present paper presents details of the latter derivation
as well as a discussion of new methods that may be useful for further
applications.

\section{Associative algebra of the standard hydrogen atom}

Modeled by a simple Coulomb potential, the hydrogen atom has the Hamiltonian
\begin{equation}
 H=\frac{1}{2m}|p|^2-\frac{\alpha}{r}
\end{equation}
with constant $\alpha$, where $|p|^2=p_x^2+p_y^2+p_z^2$ and $r^2=x^2+y^2+z^2$
in Cartesian coordinates. As operators, the position and momentum components
are subject to the basic commutation relations
\begin{equation}
[\hat{x},\hat{p}_x]=[\hat{y},\hat{p}_y]=[\hat{z},\hat{p}_z]=i\hbar\,,
\end{equation}
and they are self-adjoint. These conditions define a so-called $*$-algebra,
which, together with a quantum Hamiltonian $\hat{H}$, properties of angular
momentum, and the virial theorem, will be the only ingredient in our
derivation of spectral properties. We will not make use of operators that
represent the observables on a Hilbert space of wave functions.

An eigenvalue is a property of an observable in the algebra together with a
specific eigenstate. For a derivation of spectral properties we therefore need
a notion of states on an algebra, bypassing the introduction of wave
functions. Given a $*$-algebra ${\cal A}$, a quantum state \cite{LocalQuant}
is defined as a positive linear functional $\omega\colon{\cal A}\to {\mathbb
  C}$ from the algebra to the complex numbers, such that
$\omega(\hat{a}^*\hat{a})\geq 0$ for all $a\in{\cal A}$. In addition, a state
obeys the normalization condition $\omega(\hat{{\mathbb I}})=1$ where
$\hat{{\mathbb I}}\in{\cal A}$ is the unit.  The evaluation $\omega(\hat{a})$
is then the expectation value of $\hat{a}\in{\cal A}$, and moments such as
$\omega(\hat{a}^n)$ for integer $n$ define a probability distribution for
measurements of the observable $a$ if $\hat{a}$ is self-adjoint, $a^*=a$. Our
aim is to derive properties of eigenvalues $\lambda$ of a quantum Hamiltonian
$\hat{H}\in{\cal A}$ for hydrogen through a suitable subset the moment
conditions
\begin{equation} \label{Eigen}
 \omega(\hat{a} (\hat{H}-\lambda))=0 \quad\mbox{for all}\quad \hat{a}\in{\cal
   A}\,. 
\end{equation}
We have to find a useful subset of $\hat{a}\in{\cal A}$ in order to make this
derivation feasible.

\subsection{Subalgebra for spherical symmetry}

Instead of applying standard position and momentum components, spherical
symmetry can be used to introduce a promising subset of algebra elements. A
subalgebra of certain spherically symmetric elements of ${\cal A}$ is
generated by the three elements
\begin{equation} \label{rPQ}
 \hat{r}\quad,\quad \hat{P}=\hat{r}|\hat{p}|^2 \quad,\quad
 \hat{Q}=
\hat{x}\hat{p}_x+\hat{y}\hat{p}_y+\hat{z}\hat{p}_z-i\hbar  \,. 
\end{equation}
Linear combinations of these generators form a 3-dimensional Lie algebra with
basic relations
\begin{equation} \label{Alg}
 [\hat{r},\hat{Q}] = i\hbar \hat{r}\quad,\quad
 [\hat{r},\hat{P}]=2i\hbar\hat{Q}\quad,\quad [\hat{Q},\hat{P}]=i\hbar\hat{P}\,,
\end{equation}
isomorphic to ${\rm so}(2,1)$. (Closely related algebras have been used for
derivations of the hydrogen spectrum in deformation quantization
\cite{DefQuant2,DefQuantKepler,DefQuantHydro}. Our application of this algebra
follows different methods, and our extension to non-associative hydrogen in
the next section is completely new.) Its Casimir element is given by
\begin{equation} \label{K}
 \hat{K}:=\frac{1}{2}(\hat{r}\hat{P}+\hat{P}\hat{r})-\hat{Q}^2\,.
\end{equation}
Using the definitions (\ref{rPQ}), $\hat{K}$ turns out to equal the square of
angular momentum.

The commutators (\ref{Alg}) rely on $\hat{P}$ and $\hat{Q}$ being defined in
the specific orderings shown in their definition (\ref{rPQ}), making them not
self-adjoint. Completing the definition of a $*$-subalgebra, their adjointness
relations can be derived from the basic commutators of Cartesian position and
momentum components: In addition to $\hat{r}^*=\hat{r}$, we have
\begin{equation} \label{Qstar}
 \hat{Q}^*=\hat{Q}-i\hbar \hat{{\mathbb I}}
\end{equation}
and
\begin{equation} \label{Pstar}
 \hat{P}^*= \hat{P}-2i\hbar \hat{r}^{-1}\hat{Q}=\hat{P}-2i\hbar
 \hat{Q}\hat{r}^{-1}-2\hbar^2\hat{r}^{-1}\,. 
\end{equation}
At this point, we assume that the subalgebra generated by $\hat{r}$,
$\hat{P}$ and $\hat{Q}$ is suitably extended such that it includes an inverse
of $\hat{r}$ which, like $\hat{r}$, is also self-adjoint. The adjointness
relations imply the conditions
\begin{equation} \label{QIm}
{\rm Im}\,\omega(\hat{Q})=\frac{1}{2i}\omega(\hat{Q}-\hat{Q}^*)=\frac{1}{2}\hbar
\end{equation}
and
\begin{eqnarray}
{\rm Im}\,\omega(\hat{r}\hat{P})&=& \frac{1}{2i}
\left(\omega(\hat{r}\hat{P})-\omega(\hat{P}^*\hat{r})\right)= \frac{1}{2i}
\left(\omega(\hat{P}\hat{r}+2i\hbar\hat{Q})- \omega(\hat{P}\hat{r}-2i\hbar
  \hat{Q}-2\hbar^2\hat{{\mathbb I}})\right)\nonumber\\
&=& \hbar \omega(\hat{Q}+\hat{Q}^*)=
2\hbar{\rm Re}\,\omega(\hat{Q})
\end{eqnarray}
for expectation values in any state $\omega$ on the algebra, in addition to
${\rm Im}\,\omega(\hat{r})=0$.

In deriving (\ref{Pstar}), we have made use of the commutator
\begin{equation}
[\hat{r},|\hat{p}|^2] = 2i\hbar\hat{r}^{-1}\hat{Q}=2i\hbar 
 (\hat{Q}-i\hbar\hat{{\mathbb I}})\hat{r}^{-1}
\end{equation}
which is itself based on the commutator
\begin{equation}
  [\hat{r}^{-1},\hat{Q}] = -i\hbar \hat{r}^{-1}\label{rinvQ}
\end{equation}
in the second step. These commutators can be computed easily in a position
representation of momentum components in $\hat{p}$ and $\hat{Q}$, which then
defines the extension of our algebra to one that includes $\hat{r}^{-1}$.
Related useful commutators are
\begin{eqnarray}
  [\hat{Q},|\hat{p}|^2] &=& 2i\hbar \hat{r}^{-1}\hat{P}\\\mbox{}
 [\hat{r}^{-1},|\hat{p}|^2] &=& 2i\hbar \hat{r}^{-3}(\hat{Q}+i\hbar)\\\mbox{}
 [\hat{r}^{-1},\hat{P}] &=& 2i\hbar \hat{r}^{-2}(\hat{Q}+i\hbar)
\end{eqnarray}
and
\begin{eqnarray}
 [\hat{P}^*,\hat{P}]&= & -2i\hbar
 \hat{r}^{-1}\hat{Q}\hat{P}^*+
 2i\hbar \hat{P}(\hat{Q}-i\hbar)\hat{r}^{-1}+ 4\hbar^2
 \hat{r}^{-1}\hat{Q}(\hat{Q}-i\hbar)\hat{r}^{-1}\\\mbox{}
 [\hat{P}^*,\hat{r}] &=& [\hat{r},\hat{P}]^*=-i\hbar \hat{r}\\\mbox{}
 [\hat{P}^*,\hat{Q}] &=& -i\hbar
 \hat{P}^*=[\hat{P}^*,\hat{Q}^*]\,.
\end{eqnarray}

Adjointness relations require us to extend the algebra by an inverse of
$\hat{r}$. Nevertheless, we will see that all moments required for a
derivation of spectral properties can be derived using relations in the linear
algebra because the expectation value $\omega(\hat{r}^{-1})$ is related to
moments of polynomial expressions by the virial theorem, which states that for
any quantum Hamiltonian $\hat{H}=\frac{1}{2}m^{-1}\hat{p}^2+ \alpha \hat{r}^n$
with some integer $n$, the expectation values of kinetic and potential energy
in a {\em stationary} state $\omega$ are related by
\begin{equation} \label{Virial}
 \omega(\hat{p}^2)=nm\alpha\omega(\hat{r}^n)\,.
\end{equation}
Since all energy eigenstates are stationary, the theorem applies in our
case. In addition, the eigenvalue condition $\omega(\hat{H}-\lambda)=0$, as a
special case of (\ref{Eigen}), implies a second condition for the same energy
expectation values:
\begin{equation}
\frac{1}{2m}\omega(\hat{p}^2)+\alpha\omega(\hat{r}^n)=\lambda\,.
\end{equation}
Therefore,
\begin{equation} \label{Virial2}
 \omega(\hat{p}^2)=\frac{2nm\lambda}{n+2} \quad,\quad
\alpha\omega(\hat{r}^n)=\frac{2\lambda}{n+2}\,.
\end{equation}
For the Coulomb potential, 
\begin{equation}
 \omega(\hat{r}^{-1})= \frac{2\lambda}{\alpha}
\end{equation}
is therefore strictly determined.

The proof of the virial theorem is brief and standard, but it is useful to
display the key ingredients to demonstrate that no Hilbert-space
representation is required.  Since $\omega$ is stationary, we have
\begin{equation} \label{VirialProof}
 0=\frac{{\rm d}\omega(\hat{Q})}{{\rm d}t} =
 -i\omega([\hat{x}\hat{p}_x+\hat{y}\hat{p}_y+\hat{z}\hat{p}_z,\hat{H}]) =
 \omega(m^{-1}|\hat{p}|^2-n\alpha \hat{r}^n) 
\end{equation}
using $[\hat{p}_x,\hat{r}]=-i\widehat{\partial r/\partial x}=-i
\hat{x}\hat{r}^{-1}$.  This result proves the virial theorem not only for
standard quantum mechanics but also for non-associative systems in the
presence of magnetic monopoles: While some associativity is applied in
computing the commutator in (\ref{VirialProof}), none of the brackets
(\ref{ppp}) appear that would be modified for non-zero magnetic charge.

\section{Angular momentum}
\label{s:AngMom}

We will use the familiar eigenvalues of angular momentum squared, which equal
the eigenvalues of $\hat{K}$ defined in (\ref{K}). The usual derivation of
these eigenvalues is, to a large degree, algebraic, but it relies on
applications of ladder operators on wave functions representing
angular-momentum eigenstates. Such an application will no longer be available
once we turn to non-associative hydrogen. We therefore provide here a complete
algebraic derivation of angular-momentum eigenvalues.

The relevant algebra in this derivation is the enveloping algebra ${\cal B}$
of the Lie algebra ${\rm su}(2)$, with self-adjoint generators $\hat{J}_x$,
$\hat{J}_y$ and $\hat{J}_z$ such that
\begin{equation}
 [\hat{J}_x,\hat{J}_y]=i\hbar\hat{J}_z \quad,\quad
 [\hat{J}_y,\hat{J}_z]=i\hbar\hat{J}_x \quad,\quad
 [\hat{J}_z,\hat{J}_x]=i\hbar\hat{J}_y\,. 
\end{equation}
An angular-momentum eigenstate $\omega_{\iota,\mu}$ with eigenvalue
$\iota$ of the square of angular momentum,
$\hat{J}^2=\hat{J}_x^2+\hat{J}_y^2+\hat{J}_z^2$, and eigenvalue $\mu$ of the
$z$-component $\hat{J}_z$ is a normalized and positive linear map from ${\cal
  B}$ to the complex numbers which obeys the conditions
\begin{equation}
\omega_{\iota, \mu}(\hat{a}(\hat{J}^2 - \iota)) = 0\quad\mbox{and}\quad
\omega_{\iota,\mu}(\hat{a}(\hat{J}_z - \mu)) = 0
\end{equation}
for all $\hat{a}\in{\cal B}$. 

Although we will not apply the ladder operators $\hat{J}_{\pm}$ to wave
functions, defined as usually as
\begin{equation}
 \hat{J}_{\pm}= \hat{J}_x\pm \hat{J}_y\,,
\end{equation}
they are still useful because they obey the identity
\begin{eqnarray*}
\hat{J}_{-}^{N} \hat{J}_{+}^N &=& \hat{J}_{-}^{N-1} (\hat{J}^2 -
\hat{J}_{z}^2-\hat{J}_{z}) \hat{J}_{+}^{N-1} \\ 
&=& \hat{J}_{-}^{N-1} \hat{J}_{+}^{N-1} (\hat{J}^2 - \hat{J}_{z}^2-\hat{J}_{z}
)+ \hat{J}_{-}^{N-1}[ (\hat{J}^2 - \hat{J}_{z}^2-\hat{J}_{z}),
\hat{J}_{+}^{N-1}] \\ 
&=& \hat{J}_{-}^{N-1} \hat{J}_{+}^{N-1}\left(\hat{J}^2 - \hat{J}_{z}^2-\hat{J}_{z}
- 2 (N-1) \hat{J}_{z} - (N-1)^2 - (N-1) \right) 
\end{eqnarray*}
on ${\cal B}$ for any positive integer $N$.  Similarly,
\begin{equation}
 \hat{J}_{+}^{N} \hat{J}_{-}^{N} = \hat{J}_{+}^{N-1}
 \hat{J}_{-}^{N-1}\left(\hat{J}^2 - \hat{J}_{z}^2 + \hat{J}_{z} + 2 (N-1)
 \hat{J}_{z} - (N-1)^2 - (N-1) \right)\,.
\end{equation}
Evaluating these identities in an eigenstate, we find
\begin{eqnarray*}
\omega_{\iota,\mu}(\hat{J}_{\mp}^{N} \hat{J}_{\pm}^{N}) &=&
\omega_{\iota,\mu}(\hat{J}_{\mp}^{N-1} \hat{J}_{\pm}^{N-1}) \left(\iota - (\mu
\pm (N-1))^2 \mp (\mu \pm (N-1))\right) \\ 
&=& \prod \limits_{n = 0}^{N-1} (\iota - (\mu\pm n)^2 \mp (\mu \pm n))
\end{eqnarray*}
by iteration.

Since we have
$\hat{J}_{-}^{N} \hat{J}_{+}^{N} = (\hat{J}_{+}^{N})^* (\hat{J}_{+}^{N})$ and
$\hat{J}_{+}^{N} \hat{J}_{-}^{N} = (\hat{J}_{-}^{N})^* (\hat{J}_{-}^{N})$,
positivity of $\omega_{\iota,\mu}$ implies
\begin{equation} \label{prodN}
\prod \limits_{n = 0}^{N} (\iota - (\mu \pm n)^2 \pm (\mu \pm n)) \geq 0
\end{equation}
for all $N\geq0$.  The second term in each factor, $- (\mu \pm n)^2$, is a
negative square which can grow arbitrarily negative. Therefore, only finitely
many factors in the products (\ref{prodN}) can be non-zero, such that,
for some positive integers $n_+$ and $n_-$, we have
\begin{equation}
\iota - (\mu + n_+)^2 - (\mu + n_-) = 0 \quad \mbox{and}\quad
\iota - (\mu - n_-)^2 + (\mu - n_-) = 0 \,.
\end{equation}
Solving these two equations implies that the eigenvalues are of the form
\begin{equation}
\iota  = \left(\frac{n_- + n_+}{2}\right) \left(\frac{n_- + n_+}{2} +
  1\right) \quad\mbox{and}\quad 
\mu =  \frac{n_- - n_+}{2}\,,
\end{equation}
which can be recognized as the familiar eigenvalues in finite-dimensional
irreducible representations of ${\rm su}(2)$. 

\subsection{Eigenvalue constraint}

We proceed with our derivation of energy eigenvalues.  The Hamiltonian is not
polynomial in basic observables of the linear algebra. However, some of the
conditions (\ref{Eigen}) are defined on the linear algebra, provided $\hat{a}$
has at least one factor of $\hat{r}$ on its right. For instance, a single such
factor, $\hat{a}=\hat{r}$, replaces the non-polynomial $\hat{H}-\lambda$ in
(\ref{Eigen}) with the linear expression
\begin{equation} \label{C}
 \hat{C}:=\hat{r}(\hat{H}-\lambda)= \frac{1}{2m}\hat{P}-\lambda
 \hat{r}-\alpha\,. 
\end{equation}
A subset of the spectrum conditions (\ref{Eigen}) can therefore be written in
terms of $\hat{C}$ as the constraint equations
\begin{equation} \label{C0}
 \omega(\hat{b}\hat{C})=0 \quad\mbox{for all}\quad \hat{b}\in{\cal A}\,.
\end{equation}
These constraints might not be sufficient to obtain the full spectrum based on
(\ref{Eigen}), but any condition on eigenvalues derived from (\ref{C0}) also
applies to the full eigenvalues.

In what follows, we therefore replace the self-adjoint Hamiltonian $\hat{H}$
with a constraint operator $\hat{C}$ that, by definition, is {\em not}
self-adjoint. Dealing with constraints that are not self-adjoint requires some
care. In particular, while a self-adjoint constraint generates a gauge flow in
much the same way as a self-adjoint Hamiltonian generates time evolution,
there are additional terms in the relationship between the flow and the
commutator with $\hat{C}$ when the constraint $\hat{C}$ is not self-adjoint.

For a self-adjoint Hamiltonian $\hat{H}$, a time-dependent state $\omega_t$ by
definition evolves according to
\begin{equation}
\frac{{\rm d}\omega_t(\hat{O})}{{\rm d}t} = \frac{{\rm d}}{{\rm
     d}t} \omega_t(\exp(it \hat{H}/\hbar) \hat{O}
 \exp(-it \hat{H}/\hbar))= \frac{\omega_t([\hat{O},\hat{H}])}{i\hbar}
\end{equation}
for all $\hat{O}\in{\cal A}$.  Similarly, defining the gauge flow generated by
$\hat{C}$ through the (non-unitary) operator
$\hat{F}_{\epsilon}=\exp(-i\epsilon \hat{C}/\hbar)$ a gauge-dependent state
$\omega_{\epsilon}$ flows according to
\begin{equation}
 \frac{{\rm d}\omega_{\epsilon}(\hat{O})}{{\rm d}\epsilon} = \frac{{\rm d}}{{\rm
     d}\epsilon} \omega_{\epsilon}(\hat{F}_{\epsilon}^*\hat{O}\hat{F}_{\epsilon})
\end{equation}
because this condition implies that any state solving the constraint equation
(\ref{C0}) is preserved by the flow: We then have
\begin{equation}
 \omega(\hat{F}_{\epsilon}^*\hat{O}\hat{F}_{\epsilon})=\omega(\hat{O})
\end{equation} 
for all $\hat{O}\in{\cal A}$ if $\omega(\hat{b}\hat{C})=0$ for all
$\hat{b}\in{\cal A}$.  Infinitesimally, applying the flow operator implies a
relationship,
\begin{equation}
 \frac{{\rm d}\omega_{\epsilon}(\hat{O})}{{\rm d}\epsilon} = \frac{{\rm d}}{{\rm
     d}\epsilon} \omega_{\epsilon}(\exp(i\epsilon \hat{C}^*/\hbar) \hat{O}
 \exp(-i\epsilon \hat{C}/\hbar))=
 \frac{\omega_{\epsilon}(\hat{O}\hat{C}-\hat{C}^*\hat{O})}{i\hbar}\,,
\end{equation}
that is not directly related to the commutator of $\hat{O}$ and $\hat{C}$
because of the presence of a $\hat{C}^*$. In our specific case, we can use
\begin{equation}
\hat{C}^*=\hat{C}-\frac{i\hbar}{m} \hat{r}^{-1}\hat{Q}= \hat{C}-\frac{i\hbar}{m}
\hat{Q}\hat{r}^{-1}-\frac{\hbar^2}{m} \hat{r}^{-1}
\end{equation}
and arrive at
\begin{equation} \label{Ceps}
 \frac{{\rm d}\omega_{\epsilon}(\hat{O})}{{\rm d}\epsilon} =
 \frac{\omega_{\epsilon}([\hat{O},\hat{C}])}{i\hbar}+
 \frac{\omega_{\epsilon}(\hat{Q}\hat{r}^{-1}\hat{O})}{m}- \frac{i\hbar
 \omega_{\epsilon}(\hat{r}^{-1}\hat{O})}{m}\,.
\end{equation}

The constraint equations (\ref{C0}) play the same role as stationarity of
eigenstates. Since every energy eigenstate $\omega$ is gauge invariant under
the flow generated by $\hat{C}$, (\ref{Ceps}) implies that
\begin{equation} \label{Ceps2} 
\frac{\omega([\hat{O},\hat{C}])}{i\hbar}
  =-\frac{\omega((\hat{Q}-i\hbar \hat{{\mathbb I}})\hat{r}^{-1}\hat{O})}{m}
\end{equation}
for any such state.  Since $\hat{r}^{-1}$ appears on the right, these
equations give us another way to derive moments involving $\hat{r}^{-1}$.  For
instance, for $\hat{O}=\hat{r}$, we obtain
\begin{equation} 
  \omega({\hat{Q}})= \frac{\omega([\hat{r},\hat{P}])}{2i\hbar}=
  \frac{m\omega([\hat{r},\hat{C}])}{i\hbar}=
  -\omega(\hat{Q}-i\hbar\hat{{\mathbb I}})
\end{equation}
from the basic commutators in the first step, the definition of $\hat{C}$ in
the second step and an application of equation~(\ref{Ceps2}) in the last step.
Thus,
\begin{equation} \label{Qexp}
 \omega(\hat{Q})= \frac{1}{2}i\hbar\,,
\end{equation}
which is consistent with the reality condition (\ref{QIm}) and in
addition shows that ${\rm Re}\,\omega(\hat{Q})=0$ for stationary states.

In another example, choosing $\hat{O}=\hat{Q}$ and using the basic
commutators, as well as (\ref{rinvQ}), implies
\begin{equation} \label{omegaPr}
\frac{1}{2m}\omega(\hat{P}) + \lambda \omega(\hat{r})+
\frac{1}{m}\omega(\hat{Q}^2\hat{r}^{-1})- \frac{\hbar^2}{m}
\omega(\hat{r}^{-1})=0\,. 
\end{equation}
In this expression, The first term is given by
\begin{equation} \label{Pexp}
\frac{1}{2m}\omega(\hat{P})=\lambda\omega(\hat{r})+\alpha
\end{equation}
using the constraint, $\omega(\hat{C})=0$.  The factor of $\hat{Q}^2$ in the
third moment can be eliminated by using the Casimir $\hat{K}$, such that
\begin{eqnarray}
 \hat{Q}^2\hat{r}^{-1}&=&
 \left(-\hat{K}+\frac{1}{2}(\hat{r}\hat{P}+\hat{P}\hat{r})\right)
 \hat{r}^{-1}\nonumber\\
&=& -\hat{K}\hat{r}^{-1} +\frac{1}{2} [\hat{r},\hat{P}]\hat{r}^{-1}
+\hat{P}\nonumber\\
&=&-\hat{K}\hat{r}^{-1} +i\hbar\hat{Q}\hat{r}^{-1} +\hat{P}\,.
\end{eqnarray}
The final appearance of $\hat{Q}$ in the second term of this new expression
can be eliminated by applying (\ref{Ceps2}) to $\hat{O}=\hat{{\mathbb I}}$:
$i\hbar\omega(\hat{Q}\hat{r}^{-1}) =\hbar^2 \omega(\hat{r}^{-1})$.  Thus,
equation~(\ref{omegaPr}) implies
\begin{equation}
 3\alpha + 4\lambda  \omega(\hat{r})- \frac{K}{m} \omega(\hat{r}^{-1})=0
\end{equation}
with an eigenvalue $K=\hbar^2\ell(\ell+1)$ of $\hat{K}$ if we assume, as
usual, that our eigenstate is simultaneously one of energy and angular
momentum and then use the derivation given in Section~\ref{s:AngMom}. Using
the virial theorem to replace $\omega(\hat{r}^{-1})=2\lambda/\alpha$, we
obtain
\begin{equation}\label{rlambda}
 \omega(\hat{r}) = -\frac{1}{2}\ell(\ell+1)
 \frac{\hbar^2}{m\alpha}-\frac{3}{4}\frac{\alpha}{\lambda}
\end{equation}
for the radius expectation value.  This equation gives the correct expression
for the $r$-expectation value in all energy eigenstates in terms of the
eigenvalue $\lambda$.

\subsection{Uncertainty relations}

So far, we have not obtained any restriction on the eigenvalues $\lambda$ that
may appear in (\ref{rlambda}). Such restrictions cannot be derived by using
only the eigenmoment equation (\ref{Eigen}). In addition, we have to impose
conditions that ensure that $\omega$ is positive (or ``normalizable'' in
quantum-mechanics lingo). In order to keep the discussion more physically
intuitive, we implement positivity through the equivalent conditions implied
by uncertainty relations.

\subsubsection{General derivation}

To arrive at uncertainty relations, we follow standard results that
imply the Cauchy--Schwarz inequality
\begin{equation} \label{CS}
 \omega(\hat{a}^*\hat{a})\omega(\hat{b}^*\hat{b})\geq
 |\omega(\hat{b}^*\hat{a})|^2
\end{equation}
for all $\hat{a},\hat{b}\in {\cal A}$ and any state $\omega$. The proof
proceeds by defining a new algebra element $\hat{a}':=\hat{a}\exp(-i{\rm
  arg}\,\omega(\hat{b}^*\hat{a}))$, designed such that
$|\omega(\hat{b}^*\hat{a})|=\omega(\hat{b}^*\hat{a}')$. This intermediate step
allows us to rewrite the positivity condition as
\begin{eqnarray}
 0&\leq& \omega\left( \left(\sqrt{\omega(\hat{b}^*\hat{b})} \hat{a}'-
   \sqrt{\omega(\hat{a}^{\prime 
       *}\hat{a}')}\,\hat{b}\right)^* \left(\sqrt{\omega(\hat{b}^*\hat{b})}
   \hat{a}'- 
   \sqrt{\omega(\hat{a}^{\prime 
       *}\hat{a}')}\,\hat{b}\right)\right)\nonumber\\
&=& 2\omega(\hat{b}^*\hat{b})\omega(\hat{a}^{\prime *}\hat{a}')-
\sqrt{\omega(\hat{b}^*\hat{b})\omega(\hat{a}^{\prime 
    *}\hat{a}')} \left(\omega(\hat{a}^{\prime
    *}\hat{b})+\omega(\hat{b}^*\hat{a}')\right)\nonumber\\ 
&=& 2\omega(\hat{b}^*\hat{b})\omega(\hat{a}^{\prime *}\hat{a}')-
2\sqrt{\omega(\hat{b}^*\hat{b})\omega(\hat{a}^{\prime 
    *}\hat{a}')} |\omega(\hat{b}^*\hat{a})|
\end{eqnarray}
and to conclude that
\[
|\omega(\hat{b}^*\hat{a})|\leq \sqrt{\omega(\hat{a}^*\hat{a})}
\sqrt{\omega(\hat{a}^{\prime *}\hat{a}')}=
  \sqrt{\omega(\hat{a}^*\hat{a})\omega(\hat{b}^*\hat{b})}\,.
\]
Importantly, the result does not require associativity; see also
\cite{NonAss}. 

Choosing $\hat{a}=\hat{O}_1-\omega(\hat{O}_1)\hat{{\mathbb I}}$ and
$\hat{b}=\hat{O}_2-\omega(\hat{O}_2)\hat{{\mathbb I}}$ for self-adjoint
$\hat{O}_1$ and $\hat{O}_2$, we compute the variances
$\omega(\hat{a}^*\hat{a})=(\Delta O_1)^2$, $\omega(\hat{b}^*\hat{b})=(\Delta
O_2)^2$ of $O_1$ and $O_2$, respectively, and $\omega(\hat{b}^*\hat{a})=
\Delta(O_1O_2)+ \omega([\hat{O}_1,\hat{O}_2])$ is related to their covariance
$\Delta(O_1O_2)$. In this way, the Cauchy--Schwarz inequality implies
Heisenberg's uncertainty relation
\begin{equation}
 (\Delta O_1)^2(\Delta O_2)^2-\Delta(O_1O_2)^2\geq
 \left(\sum_IC_{12}^I\omega(\hat{O}_I)\right)^2
\end{equation}
for any pair of observales $O_1$ and $O_2$ whose algebra elements $\hat{O}_1$
and $\hat{O}_2$ are two of the generators of a linear subalgebra of ${\cal A}$
with structure constants $C_{IJ}^K$: $[\hat{O}_1,\hat{O}_2]= \sum_{I}C_{12}^I
\hat{O}_I$ with a summation range equal to the dimension of the
subalgebra. According to (\ref{Alg}), we can apply such an uncertainty
relation to any pair of the generators $(\hat{r},\hat{P},\hat{Q})$. Some of
our generators, $\hat{P}$ and $\hat{Q}$, are not self-adjoint. In such a case,
according to the derivation shown here where $(\Delta O_1)^2$ results from
$\omega(\hat{a}^*\hat{a})$, any variance of a self-adjoint expression should
be replaced with the covariance of the algebra element and its adjoint, such as
\begin{equation}
 \Delta(\bar{P}P)=\frac{1}{2}\omega(\hat{P}^*\hat{P}+\hat{P}\hat{P}^*)
-|\omega(\hat{P})|^2\,.
\end{equation}

\subsubsection{Relevant moments}

We will apply these (generalized) uncertainty relations to pairs of algebra
elements given by $\hat{r}$, $\hat{P}$ and $\hat{Q}$. An explicit evaluation
in terms of the energy eigenvalue requires a derivation of the moments
$\omega(\hat{r}^2)$, $\omega(\hat{P}^*\hat{P})$, $\omega(\hat{Q}^*\hat{Q})$,
$\omega(\hat{r}\hat{P})$, $\omega(\hat{r}\hat{Q})$, and
$\omega(\hat{Q}^*\hat{P})$. Also here, we can exploit (\ref{Ceps2}) for
different choices of $\hat{O}$, as well as (\ref{C0}) for various choices of
$\hat{b}$. 

We first compute one of the $\hat{Q}$-related moments.  First, equation
(\ref{Ceps2}) evaluated for $\hat{O}=\hat{r}^2$ implies
\begin{equation}
\frac{\omega(\hat{r}\hat{Q}+\hat{Q}\hat{r})}{m}=
\frac{\omega([\hat{r}^2,\hat{C}])}{i\hbar} =
-\frac{\omega((\hat{Q}-i\hbar\hat{{\mathbb I}}) \hat{r})}{m}=
\frac{\omega(\hat{r}\hat{Q}-2\hat{Q}\hat{r})}{m} 
\end{equation}
using the basic commutators (\ref{Alg}) in the last step. Therefore,
\begin{equation}
 \omega(\hat{Q}\hat{r})=0 \quad\mbox{and}\quad \omega(\hat{r}\hat{Q})=
 \omega([\hat{r},\hat{Q}])= i\hbar\omega(\hat{r})\,.
\end{equation}
Together with (\ref{Qexp}), we arrive at
\begin{equation} \label{DeltarQ}
 \Delta(rQ)=0\,.
\end{equation}

For the $\hat{P}$-related moments, we apply (\ref{C0}) with $\hat{b}$ equal to
the three linear generators $\hat{r}$, $\hat{Q}$ and $\hat{P}$ as well as the
adjoint $\hat{P}^*$, giving us four equations,
\begin{equation} \label{br}
\omega(\hat{r}^2)=\frac{1}{2m\lambda}\omega(\hat{r}\hat{P})-
\frac{\alpha}{\lambda} \omega(\hat{r})
\end{equation}
from $\hat{b}=\hat{r}$,
\begin{equation} \label{QP}
 \omega(\hat{Q}\hat{P})=i\hbar m\alpha
\end{equation}
from $\hat{b}=\hat{Q}$ using $\omega(\hat{Q}\hat{r})=0$ as just derived,
\begin{equation}
\omega(\hat{P}^2)=2m\lambda\omega(\hat{P}\hat{r})+
2m\alpha\omega(\hat{P})
\end{equation}
from $\hat{b}=\hat{P}$, and 
\begin{eqnarray}
 \omega(\hat{P}^*\hat{P})&=& 2m\lambda\omega(\hat{P}^*\hat{r})+ 2m\alpha
 \omega(\hat{P}^*)\nonumber\\
&=&2m\lambda\omega(\hat{P}\hat{r})+
2m\alpha\overline{\omega(\hat{P})}- 2m\lambda\hbar^2
\end{eqnarray}
from $\hat{b}=\hat{P}^*$ using (\ref{Pstar}) and (\ref{Qexp}). We apply gauge
invariance (\ref{Ceps2}) to $\hat{O}=\hat{r}\hat{Q}$, such that
\begin{equation}
 \frac{1}{2m}\omega(\hat{r}\hat{P})+ \frac{2}{m}\omega(\hat{Q}^2)
+\lambda\omega(\hat{r}^2)+\frac{1}{2m}\hbar^2=0\,.
\end{equation}

In the last equation, we replace $\hat{Q}^2$ with the square of angular
momentum and therefore $\hat{K}$, as before. 
Together with (\ref{br}) as well as (\ref{rlambda}), we obtain
\begin{equation} \label{r2lambda}
 \omega(\hat{r}^2)= \frac{3}{4}\frac{\ell(\ell+1)\hbar^2}{m\lambda}+
 \frac{5}{8} \frac{\alpha^2}{\lambda^2}- \frac{1}{4}
 \frac{\hbar^2}{m\lambda}
\end{equation}
which, like (\ref{rlambda}), is valid for all energy eigenstates in terms of
$\lambda$. The remaining equations then allow us to solve for the
$\hat{P}$-related moments
\begin{eqnarray}
 \omega(\hat{r}\hat{P}) &=& \frac{1}{2}\ell(\ell+1)\hbar^2- \frac{1}{4}
 m\frac{\alpha^2}{\lambda}- \frac{1}{2}\hbar^2 \label{rP}\\
 \omega(\hat{P}\hat{r}) &=& \frac{1}{2}\ell(\ell+1)\hbar^2- \frac{1}{4}
 m\frac{\alpha^2}{\lambda}+ \frac{1}{2}\hbar^2\label{Pr}\\
 \omega(\hat{P}^*\hat{P}) &=& -\ell(\ell+1) m\lambda\hbar^2+
 \frac{1}{2}m^2\alpha^2- m\lambda\hbar^2\,.
\end{eqnarray}

The $\hat{Q}$-related moment $\omega(\hat{Q}\hat{P})=i\hbar m\alpha$ is
already determined by (\ref{QP}), which together with (\ref{Qstar}),
(\ref{Pexp}) and (\ref{rlambda}) gives
\begin{equation}
 \omega(\hat{Q}^*\hat{P})=i\hbar\left(m\alpha- \omega(\hat{P})\right)= -i\hbar
 m \left(\alpha+ 2\lambda\omega(\hat{r})\right)= i\hbar 
 \left(\frac{1}{2}m\alpha+ 
  \frac{ \lambda \ell(\ell+1)\hbar^2}{\alpha}\right)\,,
\end{equation}
which is the second $\hat{Q}$-related moment relevant for uncertainty
relations. The final moment, $\omega(\hat{Q}^*\hat{Q})$, is related to the
square of angular momentum by
\begin{equation} \label{QbarQ}
 \omega(\hat{Q}^*\hat{Q})= \omega(\hat{Q}^2)-i\hbar \omega(\hat{Q})=
 -\omega(\hat{K})+ \frac{1}{2}\omega(\hat{r}\hat{P}+\hat{P}\hat{r})-
 i\hbar \omega(\hat{Q})= -\frac{1}{2}\ell(\ell+1)\hbar^2 -
 \frac{m\alpha^2}{4\lambda} +\frac{1}{2}\hbar^2\,,
\end{equation}
using (\ref{rP}), (\ref{Pr}) and (\ref{Qexp}). 

We are now in a position to impose positivity of $\omega$. Heisenberg's 
uncertainty relation for our variables include
\begin{equation} \label{UncertrP}
 (\Delta r)^2\Delta(\bar{P}P)\geq |\Delta(rP)+i\hbar\omega(\hat{Q})|^2
\end{equation}
which is always saturated for our solutions, without restrictions on
$\lambda$. In fact, the equality in this statement is implied by the
eigenvalue constraint (\ref{C0}), such that $\omega(\hat{b}\hat{C})=0$ and
$\omega(\hat{C}^*\hat{b})=0$ for any $\hat{b}\in{\cal A}$. Since $\hat{C}$ is
linear in $\hat{P}$ and $\hat{r}$, any $P$ in the moments in (\ref{UncertrP})
can be replaced by an $r$ as follows:
\begin{eqnarray}
 \Delta(rP)+i\hbar\omega(\hat{Q})&=&
 \frac{1}{2}\omega\left((\hat{r}-\omega(\hat{r}))(\hat{P}-
\omega(\hat{P}))+ (\hat{P}-
\omega(\hat{P}))(\hat{r}-\omega(\hat{r})) \right)+ \frac{1}{2}
\omega([\hat{r},\hat{P}])\nonumber\\
&=& 
\omega\left((\hat{r}-\omega(\hat{r}))(\hat{P}-
\omega(\hat{P}))\right)\approx 2m\lambda
\omega\left((\hat{r}-\omega(\hat{r}))^2\right)\nonumber\\
&=& 2m\lambda (\Delta r)^2
\end{eqnarray}
and
\begin{equation}
 \Delta(\bar{P}P)= \frac{1}{2}
 \omega\left((\hat{P}^*-\overline{\omega(\hat{r})})(\hat{P}- 
\omega(\hat{P}))+ (\hat{P}- 
\omega(\hat{P}))(\hat{P}^*-\overline{\omega(\hat{r})})\right) \approx
(2m\lambda)^2 (\Delta r)^2
\end{equation}
where $\approx$ indicates equality on states obeying the constraint
(\ref{C0}).

The remaining inequalities,
\begin{equation}  \label{UncertrQ}
 (\Delta r)^2 \Delta(\bar{Q}Q)\geq
|\Delta(rQ)+\frac{1}{2}i\hbar\omega(\hat{r})|^2
\end{equation}
and
\begin{equation}
 \Delta(\bar{Q}Q)\Delta(\bar{P}P)\geq
|\Delta(\bar{Q}P)+\frac{1}{2}i\hbar\omega(\hat{P})|^2\,,
\end{equation}
imply the the same condition on solutions of the constraint (\ref{C0}), but
one that non-trivially restricts the values of $\lambda$.

\subsection{Energy eigenvalues}

We evaluate the inequality (\ref{UncertrQ}) explicitly, using a simplification
implied by (\ref{DeltarQ}). For the variances on the left, we have
\begin{eqnarray}
 (\Delta r)^2&=& \omega(\hat{r}^2)-\omega(\hat{r})^2=
 \frac{3}{4}\frac{\ell(\ell+1)\hbar^2}{m\lambda}+ 
 \frac{5}{8} \frac{\alpha^2}{\lambda^2}- \frac{1}{4}
 \frac{\hbar^2}{m\lambda}- \left(\frac{1}{2}\ell(\ell+1)
 \frac{\hbar^2}{m\alpha}+\frac{3}{4}\frac{\alpha}{\lambda}\right)^2\nonumber\\
&=& -\frac{\ell^2(\ell+1)^2\hbar^4}{4m^2\alpha^2}+
\frac{\alpha^2}{16\lambda^2}- \frac{\hbar^2}{4m\lambda}
\end{eqnarray}
from (\ref{rlambda}) and (\ref{r2lambda}), and
\begin{eqnarray}
 \Delta(\bar{Q}Q)= \omega(\hat{Q}^*\hat{Q})-|\omega(Q)|^2=
 -\frac{1}{2}\ell(\ell+1)\hbar^2 - 
 \frac{m\alpha^2}{4\lambda} +\frac{1}{4}\hbar^2
\end{eqnarray}
combining (\ref{QbarQ}) and (\ref{Qexp}). Subtracting the right-hand side
$\frac{1}{4}\hbar^2 \omega(\hat{r})^2$ off (\ref{UncertrQ}), using
$\Delta(rQ)=0$ according to (\ref{DeltarQ}), we obtain the inequality
\begin{eqnarray}
&&   \frac{\ell^3(\ell+1)^3\hbar^6}{8m^2\alpha^2} +
   \frac{\ell^2(\ell+1)^2\hbar^4}{16m^2\alpha^2}
   \left(\frac{m\alpha^2}{\lambda} -2\hbar^2\right) -
   \frac{\ell(\ell+1)\hbar^2}{32 m\lambda^2} \left(
    m\alpha^2 +2\hbar^2\lambda\right)\nonumber\\
&& -\frac{m\alpha^4}{64 \lambda^3}-
  \frac{\alpha^2\hbar^2}{16\lambda^2} -\frac{\hbar^4}{16m\lambda}\geq 0\,.
\end{eqnarray}
Upon multiplication with the positive $\lambda^2$, the left-hand side is
given by $\lambda^{-1}$ times a polynomial in $\lambda$ of degree three, which
can be factorized as
\begin{equation} \label{Uncertlambda}
 \frac{(\ell+1)^2\hbar^6}{8m^2\alpha^2\lambda}
\left(\ell^2\lambda+\frac{1}{2} \frac{m\alpha^2}{\hbar^2}\right)
\left(\lambda+\frac{1}{2}
   \frac{m\alpha^2}{\hbar^2(\ell+1)^2}\right) 
 \left((\ell^2+\ell-1)\lambda-\frac{1}{2}\frac{m\alpha^2}{\hbar^2}\right)\geq
 0\,. 
\end{equation}

The central parenthesis demonstrates that the inequality is saturated for any
energy eigenvalue of the hydrogen problem with maximal angular momentum for a
given quantum number $n$, such that $\ell=n-1$, using the standard expression
\begin{equation}
 \lambda_n=-\frac{m\alpha^2}{2\hbar^2 n^2}=-\frac{m\alpha^2}{2\hbar^2
   (\ell+1)^2}\,. 
\end{equation}
Each degenerate energy level therefore contains a state that saturates an
uncertainty relation, (\ref{UncertrQ}), even if it is highly excited. This
surprising result extends an observation made in \cite{WeakMono,EigenPert} for
the harmonic oscillator to the hydrogen problem. 

\subsection{Spectral conditions from uncertainty relations}

The saturation result makes use of the known formula for energy eigenvalues of
the hydrogen problem. Keeping in mind our aim to apply algebraic methods to
the non-associative generalization of the problem in the presence of small
magnetic charges, we are interested also in an independent derivation of
spectral properties directly from the inequality (\ref{Uncertlambda}). To this
end, we first note that the left-hand side of this inequality approaches
positive infinity for $\lambda\to-\infty$, while it has negative
roots.  In order to demonstrate this result it is useful
to split the discussion into two case, $\ell=0$ and $\ell>0$. In the first
case, we can rewrite the inequality as
\begin{equation}
- \frac{\hbar^4}{16m^2 \lambda}
\left(\lambda+\frac{1}{2}
   \frac{m\alpha^2}{\hbar^2}\right)^2 \geq 0\,,
\end{equation}
which eliminates all positive $\lambda$ (where we have a continuous spectrum
and therefore no normalizable states $\omega$), and distinguishes the
ground-state energy $\lambda=-\frac{1}{2} m\alpha^2/\hbar^2$ through a
saturation condition. In the second case, the inequality written as
\begin{equation} 
 \frac{\ell^2(\ell+1)^2\hbar^6}{8m^2\alpha^2}
 \left(\lambda+\frac{1}{2} \frac{m\alpha^2}{\hbar^2\ell^2}\right)
\left(\lambda+\frac{1}{2}
   \frac{m\alpha^2}{\hbar^2(\ell+1)^2}\right) 
 \left(\ell^2+\ell-1-\frac{1}{2}\frac{m\alpha^2}{\hbar^2\lambda}\right)\geq
 0
\end{equation}
has a final parenthesis which is always positive for negative
$\lambda$. Therefore, it rules out any values of $\lambda$ between the two
roots given by the first two parentheses, 
\begin{equation}
 \lambda_1=-\frac{1}{2}
   \frac{m\alpha^2}{\hbar^2\ell^2}\quad \mbox{and}\quad
   \lambda_2=-\frac{1}{2} 
   \frac{m\alpha^2}{\hbar^2(\ell+1)^2}
\end{equation}
where $\lambda_1<\lambda_2$. All intervals between the known degenerate
eigenvalues are therefore eliminated.  (An alternative derivation of this
result not based on uncertainty relations is given in the Appendix.)

\section{Non-associative hydrogen with small magnetic charge}

A direct calculation demonstrates that the algebra generated by our basic
operators $\hat{r}$, $\hat{P}$ and $\hat{Q}$ remains unchanged if we use
monopole commutators (\ref{pp}) for the momentum components, provided the
background magnetic field obeys the condition $\vec{r}\times\vec{B}=0$. This
condition implies $\vec{B}=g(\vec{r})\vec{r}$ with some function
$g(\vec{r})$. In the static case, we need $\nabla\times\vec{B}=0$, which
is fulfilled if and only if $g(r)$ is spherically symmetric. A
monopole density $\mu(r)=\nabla\cdot\vec{B}$ then requires
\begin{equation}
 g(r)= \frac{Q_{\rm m}(r)}{4\pi r^3}
\end{equation}
with the magnetic charge
\begin{equation}
 Q_{\rm m}(r)=4\pi \int \mu(r)r^2{\rm d}r
\end{equation}
enclosed in a sphere of radius $r$. For a single monopole at $r=0$, $g(r)$
is constant, while $g(r)$ depends non-trivially on $r$ for a constant
monopole density. We will assume that $g(r)=g$ is constant, which combined
with the standard Coulomb potential implies that the hydrogen nucleus has
magnetic charge $g$.

Given monopole commutators for momentum components, the modified expressions
$\hat{L}_j'=\hat{L}_j+eg\hat{x_j}\hat{r}^{-1}$ satisfy the usual commutators
of angular momentum \cite{MonopoleAngMom,MagneticCharge} and therefore have
the familiar spectrum. The Casimir of the algebra generated by $\hat{r}$,
$\hat{P}$ and $\hat{Q}$ is still equal to $\hat{K}=\hat{L}{}^2$, but in terms
of the modified angular momentum, whose eigenvalues we know as derived in
Section~\ref{s:AngMom} from the standard commutators, it has an extra term:
\begin{equation} \label{KL}
 \hat{K}=\hat{L}^2=\hat{L}'{}^2- e^2g^2 \hat{{\mathbb
     I}}\,. 
\end{equation}
(For a monopole density with non-constant $g$, $\hat{K}$ and
$\hat{L}'{}^2$ cannot be diagonalized simultaneously and an
independent method would have to be used to find eigenvalues of $\hat{K}$.)

For a single monopole at the center, the spectrum of $\hat{K}$, according to
(\ref{KL}) has a simple constant shift compared with the spectrum of
$\hat{L}'{}^2$, which is known to break the degeneracy of the energy spectrum
for magnetic monopoles that obey Dirac's quantization condition
\cite{MonopoleHydro}. This condition, $eg=\frac{1}{2}\hbar$, implies a large
value of the smallest non-zero magnetic charge because the electric fine
structure constant is small. Dirac monopoles in a hydrogen nucleus would
therefore be large perturbations that strongly modify the energy
spectrum. They can easily be ruled out by standard spectroscopy.  Dirac's
quantization condition can be violated in non-associative quantum
mechanics. Magnetic charges can then be small and might modify the energy
spectrum sufficiently weakly to be phenomenologically viable. However, a
derivation of eigenvalues in the non-associative setting remained impossible
for decades. Our methods from the preceding section can now be applied to this
question.

We will focus on a range of small magnetic charges $g$ characterized by the
condition $0< eg/\hbar < \frac{1}{2}$. As already noted, the commutators
(\ref{Alg}), the virial theorem and the Cauchy--Schwarz inequality all hold
for a non-associative monopole algebra. The only assumption that need be
modified in our previous derivation of uncertainty relations is the spectrum
of $\hat{K}$, which is no longer equal to the square of angular momentum but
instead has the eigenvalues
\begin{equation} \label{Kell}
 K_{\ell}=\ell(\ell+1)\hbar^2-e^2g^2 \,.
\end{equation}
It is convenient to parameterize the shift by replacing $\ell$ with a
non-integer quantum number
\begin{equation} 
 \tilde{\ell}=\sqrt{\left(\ell+\frac{1}{2}\right)^2-
\frac{e^2g^2}{\hbar^2}}-\frac{1}{2}\,.
\end{equation}
Substituting $\tilde{\ell}$ for $\ell$ in (\ref{Uncertlambda}) then gives us
conditions on energy eigenvalues of non-associative hydrogen. (Saturation
conditions indeed give us correct eigenvalues according to
\cite{MonopoleHydro}, but since the usual degeneracy is broken, they do not
give us the full spectrum.)

The range of $\ell$ is bounded by the fact that $\hat{K}$ is a positive
operator, such that the eigenvalues (\ref{Kell}) cannot be negative. This
condition rules out the quantum number $\ell=0$, but for small magnetic
charges the next possible value, $\ell=1/2$, is allowed. We will assume this
value for the ground state because (\ref{Uncertlambda}) tells us that the
smallest root of this equation is proportional to $-1/\ell^2$. The minimum
energy eigenvalue is therefore obtained for the smallest possible $\ell$. This
value of $\ell$ implies
\begin{equation} \label{tildel}
 \tilde{\ell}=\sqrt{1-\frac{e^2g^2}{\hbar^2}}-\frac{1}{2}
\end{equation}
which lies in the range
\begin{equation} \label{range}
\frac{1}{2}(\sqrt{3}-1)<
\tilde{\ell}<\frac{1}{2}\,.
\end{equation}
Since $\tilde{\ell}=0$ is not possible, the uncertainty relation always
rules out a range of energy eigenvalues between
\begin{equation}
 \lambda_1=-\frac{1}{2}\frac{m\alpha^2}{\hbar^2\tilde{\ell}^2}
\end{equation}
and
\begin{equation}
 \lambda_2= -\frac{1}{2}\frac{m\alpha^2}{\hbar^2(\tilde{\ell}+1)^2}\,.
\end{equation}
For any $\tilde{\ell}$ in the range (\ref{range}), $\tilde{\ell}<1$ while
$\tilde{\ell}+1>1$. Therefore, a certain non-empty range around the usual
hydrogen ground-state energy $-\frac{1}{2}m\alpha^2/\hbar^2$ is ruled out for
any value of a small magnetic charge. We conclude that even a small magnetic
charge would strongly modify the usual hydrogen spectrum and be incompatible
with spectroscopic data. This strict exclusion is possible because the
positivity of $\hat{K}$ implies a discontinuity of energy eigenvalues as
functions of the magnetic charge $g$ at $g=0$.

\section{Conclusions}

Our derivations have produced the first results about spectral properties in a
system of non-associative quantum mechanics. In particular, we have been able
to demonstrate a discontinuity in the ground-state energy of hydrogen as a
function of the magnetic charge of the nucleus. Addressing this question
requires a continuous range of the magnetic charge around zero, which cannot
be modeled by an associative treatment with Dirac monopoles for which the
magnetic charge is quantized. Non-associative quantum mechanics is able to
describe fractional magnetic charges of any value and is therefore a suitable
setting for our question.

A Hilbert-space representation of an algebra by operators acting on wave
functions is by necessity associative because for any $\psi$ in the Hilbert
space and operators $\hat{A}$, $\hat{B}$ and $C$ we have
\begin{equation}
(\hat{A}\hat{B})\hat{C}\psi=
\hat{A}\hat{B}\psi'= \hat{A}(\hat{B}\psi')=\hat{A}(\hat{B}\hat{C})\psi\,,
\end{equation}
defining $\psi'=\hat{C}\psi$ in an intermediate step. Non-associative quantum
mechanics can therefore not be represented on a Hilbert space, necessitating a
purely algebraic derivation of properties of expectation values, moments, and
eigenvalues. That such an algebraic treatment can indeed be used to derive a
complete spectrum is demonstrated in \cite{WeakMono,EigenPert}, in this case
for the (associative) harmonic oscillator as a proof of principle.  The
algebraic treatment relies on uncertainty relations in order to impose
positivity of states, replacing the more common normalizability conditions of
Hilbert-space treatments. The new methods are therefore well-suited to finding
unexpected saturation properties of eigenstates, even excited ones. As a new
result of \cite{WeakMono,EigenPert}, every eigenstate of the harmonic
oscillator saturates a suitable uncertainty relation. Saturation results even
extend to eigenstates of anharmonic systems in perturbative treatments.

Our application of related methods to non-associative hydrogen in the present
paper have not resulted yet in a full energy spectrum because we focused on
the ground state, deriving only one uncertainty relation
explicitly. Nevertheless, a saturation result has been found for this state,
indicating that the behavior seen in harmonic models might be extendable
also to excited states of hydrogen. However, the dynamical algebra of hydrogen
is more involved than the canonical algebra applicable to the harmonic
oscillator, making a generic treatment of saturation results for hydrogen more
complicated.

Our extension to non-associative hydrogen relied on several fortuitous
algebraic properties of standard hydrogen that are not affected by introducing
non-associativity of monopole type, given by a commutator (\ref{pp}) of
kinematical momentum components with a magnetic field generated by a pointlike
magnetic charge. For other non-associative algebras, or even a monopole
algebra with a continuous magnetic charge distribution, the eigenvalue problem
cannot yet be solved, presenting a challenging mathematical problem.

Our specific physical result demonstrates that the pursuit of these
mathematical questions is worthwhile. We have found that the ground-state
energy of hydrogen with a small magnetic nuclear charge $g$ is significantly
displaced from the usual value due to a discontinuity, even for
infinitesimally small magnetic charge. Spectroscopy is therefore very
sensitive to introducing a magnetic charge. In order to produce an upper bound
on $g$ consistent with observational data, we may, following \cite{WeakMono},
wash out the discontinuity implied by positivity of the non-associative
angular momentum $K$ because the eigenvalues of angular momentum squared are
determined only within some $\delta L^2$ from a purely phenomenological
viewpoint. In addition, a fundamental uncertainty in angular momentum could
also be caused by an extended magnetic charge distribution in the nucleus,
which would imply that $\hat{K}$ and $\hat{L}'^2$ no longer commute.

As an estimate of this uncertainty, we may use the value $5\cdot 10^{-19}$
given as the accuracy of recent atomic clocks \cite{LatticeClock}, which rely
on sharp spectral lines that would be affected by the same uncertainty $\delta
L^2$ if angular momentum is not sharp. The inequality $K\geq0$ for eigenvalues
of $\hat{K}$, which must always hold because $\hat{K}$ is defined as a
positive operator, then implies an upper bound
\begin{equation}
 g\leq \frac{4\pi \epsilon_0\sqrt{\delta L^2}c^2}{e}\approx 4.7\cdot 10^{-18}{\rm
   Am}= 1.4 \cdot 10^{-9} g_{{\rm Dirac}}
\end{equation}
for the magnetic charge, written here in SI units. This upper bound is a small
fraction of $g_{{\rm Dirac}}$, the smallest non-zero magnetic charge allowed
by Dirac's quantization condition in an associative treatment.

Magnetic charges of elementary particles have been bounded by various
means. Using the proton as an example, interpreted here as the nucleus of
hydrogen, our bound is not as strong as those found based on the total
magnetic charge of a large number of nucleons in macroscopic objects
\cite{MagneticChargeBound,MagneticChargeProtonNeutron}. The large number of
nucleons in macroscopic objects implies a strong magnification factor in the
latter studies if their magnetic charges add up. However, this method is not
available for those elementary particles that cannot be combined in stable
macroscopic objects, such as unstable particles or antimatter. Some of them
can nevertheless be used as substitutes of the nuclear proton in hydrogen-like
atoms, with precision spectroscopic data being available in some cases such as
muonium \cite{MuoniumSpec} or antihydrogen \cite{AntiH,AntiH2}. For instance,
muonium spectroscopy with a current accuracy of about $10^{-9}$ gives us an
upper bound on the muon's magnetic charge of $g_{\rm muon}\leq4.5\cdot
10^{-5}g_{{\rm Dirac}}$, which is better than available upper bounds based on
other methods.

\section*{Acknowledgements}

This work was supported in part by NSF grant PHY-1912168. SB is supported in
part by a McGill Space Institute fellowship and a CITA National fellowship.

\begin{appendix}

\section{Algebraic derivation of the associative hydrogen spectrum}

It is instructive to derive the standard energy spectrum of an electric charge
in a Coulomb potential by algebraic means, using the same subalgebra of
observables generated by (\ref{rPQ}) as employed in the main text but imposing
positivity of states not through uncertainty relations but, more indirectly,
through convergence properties of certain expectation values expressed as
power series. This derivation more closely resembles the standard derivation
based on convergence properties of norms of wave functions, but it is still
fully algebraic. However, it does not give rise to new saturation conditions
of uncertainty relations, and it is more difficult to extend it to
non-associative systems.

In addition to the basic commutators (\ref{Alg}), we will make use of
\begin{equation}
[\hat{r},\hat{C}] = \frac{i \hbar}{m}Q \quad\mbox{and}\quad
[\hat{Q},\hat{C}] = i \hbar \left(\frac{1}{2m}\hat{P}+\lambda \hat{r}\right)
\end{equation}
with the constraint $\hat{C}$ defined in (\ref{C}), as well as the
expectation-value equation
\begin{equation} \label{bP}
\omega( \hat{b} \hat{P}) = 2m \omega\left(\hat{b} (\lambda \hat{r} +
  \alpha)\right) 
\end{equation}
for any $\hat{b}\in{\cal A}$, implied by the eigenvalue constraint (\ref{C0}).
We will apply the invariance condition (\ref{Ceps2}) in various ways, and
use the operator (\ref{K}) in the form
\begin{equation}
\hat{K} = \hat{r}\hat{P} - i \hbar \hat{Q} - \hat{Q}^2\,.
\end{equation}

\subsection{Kramer's relation}

Our first step is the algebraic derivation of a recurrence relation for
expectation values of integer powers of $\hat{r}$ in energy eigenstates of
hydrogen, known as Kramer's relation. To this end, we derive the commutators
\begin{equation}
[\hat{r}^n,\hat{Q}] = i \hbar n \hat{r}^n \quad,\quad
[\hat{r}^n, \hat{P}] = 2 i n \hbar \hat{r}^{n-1} \hat{Q} +\hbar^2 n(n-1)
\hat{r}^{n-1} 
\end{equation}
for integer $n$, using induction and being careful with taking commutators of
powers because $[\hat{a},[\hat{a},\hat{b}]] = 0$ does not always hold for
$\hat{a},\hat{b}\in {\cal A}$.

Second, invariance applied to $\hat{O}= m \hat{r}^s$ takes the form
\begin{eqnarray*}
0 &=& \frac{m}{i \hbar} \omega([\hat{r}^s, \hat{C}] + (\hat{Q}- i \hbar)
\hat{r}^{s-1}) \\ 
&=& \frac{1}{2 i \hbar} \omega([\hat{r}^s,\hat{P}] + [(\hat{Q}-i \hbar),
\hat{r}^{s-1}]+ \hat{r}^{s-1} (\hat{Q}-i \hbar)) \\ 
&=& s\omega(\hat{r}^{s-1} \hat{Q}) - \frac{i \hbar}{2} s(s-1)
\omega(\hat{r}^{n-1}) -(s-1) i \hbar  \omega(\hat{r}^{s-1})+
\omega(\hat{r}^{s-1} (\hat{Q}-i \hbar))\\ 
&=& \frac{s+1}{2} \omega(\hat{r}^{s-1} (2 \hat{Q} - i s \hbar))\,,
\end{eqnarray*}
such that
\begin{equation} \label{omegarQ}
 \omega(\hat{r}^{s-1} \hat{Q})= \frac{1}{2} i\hbar s \omega(\hat{r}^{s-1})\,.
\end{equation}
Using this result, invariance applied to $\hat{O}= m \hat{r}^s \hat{Q}$ leads
to
\begin{eqnarray}
0 &=& \frac{m}{i \hbar} \omega([\hat{r}^s \hat{Q}, \hat{C}] + (\hat{Q}- i
\hbar) \hat{r}^{s-1} \hat{Q} ) \nonumber\\ 
&=& \frac{1}{2 i \hbar} \omega([\hat{r}^s, \hat{P}] \hat{Q}) + \frac{m}{i
  \hbar} \omega(\hat{r}^s [\hat{Q}, \hat{C}]) + \omega([(\hat{Q}- i \hbar),
\hat{r}^{s-1}] \hat{Q}) + \omega(\hat{r}^{s-1}(\hat{Q}- i \hbar) \hat{Q})
\nonumber \\ 
&=& s  \omega(\hat{r}^{s-1} \hat{Q}^2) - \frac{i \hbar}{2} s(s-1)
\omega(\hat{r}^{s-1} \hat{Q}) +  \frac{1}{2}\omega(\hat{r}^s (\hat{P}+2m
\lambda \hat{r})) - i \hbar (s-1)\omega(\hat{r}^{s-1} \hat{Q})\nonumber\\
&& +
\omega(\hat{r}^{s-1}(\hat{Q}- i \hbar) \hat{Q})  \nonumber\\
& =& (s+1) \omega(\hat{r}^{s-1} \hat{Q}^2) + \frac{1}{2} \omega(\hat{r}^s
\hat{P})+ m \lambda \omega(\hat{r}^{s+1}) - i \hbar \frac{s(s+1)}{2}
\omega(\hat{r}^{s-1} \hat{Q})  \nonumber\\
&=&  -(s+1) \omega(\hat{K} \hat{r}^{s-1}) + (s+3/2) \omega(\hat{r}^s
\hat{P})+ m \lambda \omega(\hat{r}^{s+1}) - i \hbar\frac{(s+2)(s+1)}{2}
\omega(\hat{r}^{s-1} \hat{Q})  \nonumber\\
&=&  -(s+1) \omega(\hat{K} \hat{r}^{s-1}) + (s+3/2) \omega(\hat{r}^s
\hat{P})+ m \lambda \omega(\hat{r}^{s+1}) +  \hbar^2\frac{(s+2)(s+1)s}{4}
\omega(\hat{r}^{s-1})\,.  \label{KramerK}
\end{eqnarray}
Equation~(\ref{bP}) then implies Kramer's relation
\begin{equation} \label{Kramer}
0 =  \hbar^2 (s+1) \left(\frac{s(s+2)}{4}- \ell (\ell+1)\right)
\omega(\hat{r}^{s-1}) 
+ (2 s + 3 ) m \alpha \omega( \hat{r}^{s})+2 (s+2) m \lambda \omega(
\hat{r}^{s+1} ) 
\end{equation}
after inserting the standard angular-momentum eigenvalues of $\hat{K}$. 
Incidentally, invariance applied to $\hat{O}=\hat{r}^s\hat{P}$ results in an
identity: 
\begin{eqnarray*}
&& \frac{m}{i \hbar} \omega([\hat{r}^s \hat{P}, \hat{C}]) +
\omega((\hat{Q}- i \hbar) \hat{r}^{s-1} \hat{P})  \\ 
&=& \frac{1}{2 i \hbar} \omega([\hat{r}^s, \hat{P}] \hat{P}) + \frac{m}{i
  \hbar} \omega(\hat{r}^s [\hat{P}, \hat{C}]) + \omega([(\hat{Q}- i \hbar),
\hat{r}^{s-1}] \hat{P}) + \omega(\hat{r}^{s-1}(\hat{Q}- i \hbar) \hat{P})  \\
&=& 2m(s+1)  \omega(\hat{r}^{s-1} \hat{Q}(\lambda \hat{r}+ \alpha))  - i
  \hbar m s(s+1) \omega(\hat{r}^{s-1}  (\lambda \hat{r}+ \alpha)) +
  \omega(\hat{r}^s (2 m \lambda \hat{Q})) \\ 
&=&  2m \alpha(s+1)  \omega(\hat{r}^{s-1} \hat{Q}) + 2m\lambda (s+1)
\omega(\hat{r}^{s} (\hat{Q}-i \hbar))  - i \hbar m s(s+1)
\lambda\omega(\hat{r}^{s}) \\
&& - i \hbar m\alpha s(s+1) \omega(\hat{r}^{s-1}) +
2m\lambda  \omega(\hat{r}^s\hat{Q}) =0
\end{eqnarray*}
upon using (\ref{omegarQ}).

\subsection{Spectrum}

Equipped with Kramer's relation, which we first shift down by one unit in $s$,
\begin{equation}
0 =  \hbar^2 s \left(\frac{s^2-1}{4}- \ell(\ell+1)\right) \omega(\hat{r}^{s-2}) +
(2 s + 1 ) m \alpha \omega(\hat{r}^{s-1})+2 (s+1) m \lambda \omega(
\hat{r}^{s})\,, 
\end{equation}
we can now set up a new recurrence relation. We first generalize Kramer's
relation to
\begin{equation} \label{KramerDiff}
0 =\frac{\hbar^2}{4} \omega\left((\hat{r} f(\hat{r}))'''\right) - \hbar^2
\ell(\ell+1) 
\omega(\hat{r}^{-1} f'(\hat{r})) + m \alpha 
\omega(2f'(\hat{r})+\hat{r}^{-1}f(\hat{r})) +2 m \lambda \omega((\hat{r}
f(\hat{r}))')
\end{equation}
for any analytic function $f$, where derivatives of analytic functions of
$\hat{r}$ are interpreted in the sense of formal power series.

Specializing $f(\hat{r})$ to $f_{s,k}(\hat{r}) = \hat{r}^s e^{- k \hat{r}}$
and defining 
\begin{equation} \label{kappa} 
\kappa_s(k,\lambda)= \omega(\hat{r}^s
  e^{-k\hat{r}})
\end{equation}
then gives
\begin{eqnarray*}
  0 &=& \hbar^2 s (-1 - 4 \ell (1 + \ell) + s^2) \kappa_{s-2}(k,\lambda) +
  (h^2 k (4\ell (1 + \ell) - 3 s (1 + s)) + 4 m (1 + 2 s) \alpha)
  \kappa_{s-1}(k,\lambda)\\  
  && + (3 \hbar^2 k^2 (1 + s) + 8 m (\lambda (1 + s) - k \alpha)) \kappa_s (k, \lambda) -   k (8 m \lambda+ k^2 \hbar^2) \kappa_{s+1}(k,\lambda)\,.
\end{eqnarray*}
Again shifting $s$ by defining $L_{s}(k,\lambda) = \kappa_{s-2}(k,\lambda)$,
we rewrite the previous relation as the third-order linear differential equation
\begin{eqnarray} \label{Lsrecur}
0 &=& \big( \hbar^2 s (-1 - 4 \ell (1 + \ell) + s^2) -  (h^2 k (4\ell  (1 +
\ell) - 3 s (1 + s)) + 4 m (1 + 2 s) \alpha)\partial_k 
\\  
  && + (3 \hbar^2 k^2 (1 + s) + 8 m (\lambda (1 + s) - k \alpha))\partial^2_k
  + k (8 m \lambda+ k^2 \hbar^2) \partial^3_k  
\big) L_s(k,\lambda)\,. \nonumber
\end{eqnarray}

Since our $f_{s,k}(\hat{r})$ is a bounded operator for $k>0$ and $s\geq 0$
with $\lim_{k\to\infty}f(\hat{r})=\hat{0}$, any state should be such that
$L_{s}(k,\lambda)$ is well-defined for all $k> 0$ and $s\geq0$ with
$\lim_{k\to\infty}L_s(k,\lambda)=0$ for all $\lambda$. We also know that
$L_{s}(k,\lambda)$ is well-defined for energy eigenstates at $k=0$ as long as
$s\geq 0$ is integer, because Kramer's relation together with the virial
theorem provides finite numbers for expectation values of positive integer
powers of $\hat{r}$. Under these conditions, we can perform a
Laplace-like transformation and write 
\begin{eqnarray}
L_{s}(k,\lambda) &=& \int_{0}^{\infty} a_{s,\lambda}(b,d)
(k+d(s,\lambda))^{-b} \mathrm{d}b \nonumber\\ 
&=& \sum_{n=0}^{\infty} \int_0^{1} a_{s,\lambda}(b+n,d) (k+d(s,\lambda))^{-n-b}
\mathrm{d}b \,. \label{Ls}
\end{eqnarray}
In the first line, $a_{s,\lambda}(b,d) $ may be seen as the inverse Laplace
transform of $L_{s}(e^t-d(s,\lambda))$ with respect to $t$. As we will see, it
is convenient to introduce a free displacement $d(s,\lambda)$ on which the
coefficients $a_{n,\lambda}$ will in general depend.

For further convenience, we now drop the explit dependences on $s$ and
$\lambda$ from our notation. Comparing coefficients of the expansion
(\ref{Ls}) inserted in (\ref{Lsrecur}), we obtain the recurrence relation
\begin{equation} \label{anrecur}
C_3 a(b+n-3)+C_2a(b+n-2)+C_1a(b+n-1)+C_0a(b+n)=0
\end{equation}
with
\begin{eqnarray}
C_3&=&  d (b + n-3) (b + n-2) (b + n-1) (d^2 \hbar^2 + 8 m \lambda)\\
C_2&=& (b + n-2) (b +     n-1) \left(-3 d^2 \hbar^2 (b + n-1 - s) + 8 d m
\alpha - 8 m (b + n-1 - s) \lambda\right)\\
C_1&=& (b +  n-1) \Biggl(3 d \hbar^2 (b+ n) (b + n+1)
+  d \hbar^2 \Bigl(-4 \ell (1 + \ell) + 3 s (1 + s)\Bigr)  \nonumber\\
&&+(b + n) \Bigl(-6 d \hbar^2 (1 + s) -
8 m \alpha\Bigr) + 4 m \alpha (1 + 2 s)\Biggr) \\
C_0 &=&-  h^2 (b + n - s) ((b + n - s)^2 -
(2\ell+1)^2)\,. \label{C0}
\end{eqnarray}

By definition, the support of $a$ as a function of $b$ is bounded from
below. If for a given solution $n_{\rm min}$ is the smallest integer such that
$a(b+n_{\rm min})\not=0$ while $a(b+n)=0$ for $n<n_{\rm min}$, the expression
(\ref{C0}) shows that $n_{\rm min}+b-s=0$ or $|n_{\rm
  min}+b-s|-|2\ell+1|=0$. Using the fact that $\ell$ is an integer (since we
are for now assuming the absence of a magnetic charge), $b$ must be an
integer. This result shows that $L_s(k,\lambda)$ allows an expansion as a
Laurent series of the form
\begin{equation} \label{LA}
L_{s}(k,\lambda) = \sum_{n=0}^{\infty} A_{s,\lambda}(n)
(k+d(s,\lambda))^{-n}\,.
\end{equation}
(The original coefficients $a_{s,\lambda}(b,d)$ introduced in (\ref{Ls}) are
proportional to a Dirac comb of delta functions of $b$ supported on the
integers.)

The recurrence relation for $A_{s,\lambda}(n)$ can easily be obtained from
(\ref{anrecur}) by absorbing $b$ in $n$, ignoring the shift by $b$. The
relation can be simplified further by making the choice $d = \sqrt{- 8 m
    \lambda}/\hbar$ for a given $\lambda$, such that the lowest-order term
(at order $n-3$) drops out of the recurrence. We also choose
$s=2 \ell +2$ and obtain
\begin{eqnarray}\label{an}
  0 &=&2 d (n-2) \left(d \hbar^2 (3 + 2 \ell - n) + 4 m \alpha\right) A(n-2)\\
&& + \left(d \hbar^2
  \left(8 \ell^2 + \ell (26 - 12 n) + 3 (n-3) (n-2)\right) + 4 m \alpha
(5 + 4 \ell -  2 n)\right) A(n-1) \nonumber\\
&& -  \hbar^2 (n-3 - 4 \ell) (n-2 - 2 l) A(n) \nonumber
\end{eqnarray}
after factoring out $b+n-1$.
For very large $n$ of either sign, this recursion takes the form $A(n) - 3 d
A(n-1)+ 2 d^2 A(n-2) = 0$, such that any non-zero asymptotic $A_n$ behaves
either as $d^n$ or $(2 d)^n$. However, these options would introduce a pole
for $L_{s}(k,\lambda)$, either at $k=0$ or $k=d>0$, which cannot happen for
well-defined states. Therefore, only finitely many $A(n)$ can be non-zero.
According to the $A(n)$-term in (\ref{an}), there is an $N_1$ such that
$A(n)=0$ for $n<N_1$ because $\ell$ is an integer.

For the range of $n$ where $A(n)\not=0$ to be bounded from above, the first
coefficient in (\ref{an}) the latter condition requires
\begin{equation}
 d = \frac{4 m \alpha}{\hbar^2 \nu}
\end{equation}
with some positive integer $\nu$.  Inserting this expression, we obtain
\begin{eqnarray} \label{cn}
0 &=&  2 (n+2 \ell) (n -1 - \nu) c_{n-2}\\
&& - \left( n (3n -3 - 2 \nu)+\nu-4 \ell (1 + \ell)\right)
c_{n-1}\nonumber\\
&&+n(n-1 - 2 \ell) c_n \nonumber
\end{eqnarray}
where 
\begin{equation} \label{cA}
 c_n = d^{-n} A_{n+2\ell+2}\,.
\end{equation}

There is one final condition: as all these sequences are linear with
recurrence relations that have integer coefficients (since $\ell$ is known to
be an integer) we infer that, up to $n$-independent rescalings, for a given
solution all the coefficients $c_n$ are integer multiples of the same basic
quantity, $\gamma$. Dividing the recurrence relation by $\gamma$, we have $0 =
\nu c_{n-1}/\gamma \texttt{ mod 2}$ for all $n$, because only a single term in
the coefficients of (\ref{cn}) is not guaranteed to be even. As an overall
factor of two could be absorbed into the definition of $\gamma$ (and therefore
$c_{n-1}/\gamma$ may well be odd), we conclude that $\nu = 2 N$, giving
\begin{equation}
 \delta = \frac{2 m \alpha}{\hbar^2  N}
\end{equation} 
and 
\begin{equation}
 \lambda = -\frac{m \alpha}{2 \hbar^2 N}\,,
\end{equation}
which is the known energy spectrum of hydrogen.

It is instructive to look at the detailed recurrence for the case of $\ell=0$,
which includes the ground state, such that $s=2$.  For $n=0$ in (\ref{cn}), we
obtain $c_{-1}=0$. Choosing  $n=1$ in
(\ref{cn}) then shows that $c_0=0$. For $n=2$, we obtain a non-trivial
relation that determines $c_2$ in terms of a free $c_1$: 
\begin{equation}
 c_2=3(1-\nu/2)c_1\,.
\end{equation}
For $\nu=2$, the smallest allowed value, $c_2=0$, which then implies $c_3=0$
at $n=3$. With two successive vanishing $c_n$, all the following $c_n$ are
zero.  Since $c_1$ may be non-zero, there is a non-zero solution, as required
for a non-zero expectation value of the positive operator
$\hat{r}^2e^{-k\hat{r}}$. A non-zero $c_1$ implies through (\ref{cA}) that
$A_3$ is the only non-zero coefficient, such that
\begin{equation} \label{Lsol}
 L_2(k,\lambda_0)\propto \left(k+\frac{2m\alpha}{\hbar^2}\right)^{-3}
\end{equation}
using (\ref{LA}). According to its definition (\ref{kappa}) as an expectation
value, $L_2(k,\lambda_0)=\kappa_0(k,\lambda_0)=\omega_0(e^{-k\hat{r}})$ should
be the ground-state expectation value of $e^{-k\hat{r}}$, which can easily be
confirmed to be of the form (\ref{Lsol}) using the known ground-state wave
function $\psi_0(r)\propto e^{-r/a}$ with the Bohr radius
$a=\hbar^2/(m\alpha)$.

\section{Generalization to hydrogen with a magnetic nuclear charge}

Since most of the identities used in our new derivation of Kramer's relation
hold true in the non-associative case with a pointlike magnetic monopole at
the center, we can easily generalize this relation. We only have to adjust the
spectrum of $\hat{K}$ using (\ref{Kell}) in (\ref{KramerK}) and obtain
\begin{equation} 
0 =  \hbar^2 (s+1) \left(\frac{s(s+2)}{4}- \ell (\ell+1)+e^2g^2/\hbar^2\right)
\omega(\hat{r}^{s-1}) 
+ (2 s + 3 ) m \alpha \omega( \hat{r}^{s})+2 (s+2) m \lambda \omega(
\hat{r}^{s+1} ) 
\end{equation}
as a generalization of (\ref{Kramer}).

This equation takes the form
\begin{equation} \label{KramerDiff}
0 =\frac{\hbar^2}{4} \omega\left((\hat{r} f(\hat{r}))'''\right) - \hbar^2
\left(\ell(\ell+1)-e^2g^2\right) 
\omega(\hat{r}^{-1} f'(\hat{r})) + m \alpha 
\omega(2f'(\hat{r})+\hat{r}^{-1}f(\hat{r})) +2 m \lambda \omega((\hat{r}
f(\hat{r}))')
\end{equation}
as a differential equation replacing (\ref{KramerDiff}), which in turn
implies the equation
\begin{eqnarray*}
0 &=& \Bigl(\hbar^2 s (s^2 -1- 4(\ell(\ell+1)-e^2g^2/\hbar^2))\\
&& - (4 m \alpha (2 s+
1) + k (4 
\ell(\ell+1)-4e^2g^2/\hbar^2 - 3 s(s+1)))\partial_{k}  \\ 
&&  + (8 m (s+1) - 8 k m \alpha + 3 k^2 (1+s)\hbar^2) \partial^2_k + k(8 m
\lambda+ 
k^2) \hbar^2 \partial^3_k\Bigr) L_{s}(k,\lambda) 
\end{eqnarray*}
instead of (\ref{Lsrecur}).

The recurrence relation (\ref{anrecur}) still holds with the same $C_3$ and
$C_2$, while $C_1$ and $C_0$ are replaced by

\begin{eqnarray*}
C_1'&=& (b +  n-1) \Biggl(3 d \hbar^2 (b+ n) (b + n+1)
+  d \hbar^2 \Bigl(-4 \ell (1 + \ell)+-4e^2g^2/\hbar^2 + 3 s (1 + s)\Bigr)
\nonumber\\ 
&&+(b + n) \Bigl(-6 d \hbar^2 (1 + s) -
8 m \alpha\Bigr) + 4 m \alpha (1 + 2 s)\Biggr) \\
C_0' &=&-  h^2 (b + n - s) ((b + n - s)^2 -
(2\ell+1)^2+4e^2g^2/\hbar^2)\,. \label{C0p}
\end{eqnarray*}
The same choice $d=\sqrt{-8m\lambda}/\hbar$ as in the derivation of
(\ref{an}) can be used to reduce the equation to second order, and it has the
same large-$n$ behavior as before. The sequence of $a_n$ therefore still has
only finitely many non-zero elements, which is again the case if $b-s$ is an
integer because the coefficient $b + n - s$ in the last term of the recurrence
relation has not changed. However, there is now a second possibility if $b$
and $s$ are such that $(b+n-s)^2= (2\ell+1)^2-4e^2g^2/\hbar^2$ for some
integer $n$. This condition can provide new solutions and a more complicated
spectrum.

The last coefficient, $(b+n-s)^2- (2\ell+1)^2+4e^2g^2/\hbar^2$, no longer
factorizes. Setting $b=0$ as before, we therefore obtain a relation,
\begin{eqnarray*}
&&0 =2d(n-2) (n-1) \biggl(-4  m \alpha +    d (-1 + n - s)
\hbar^2 \biggr) a_{n-2}\\ 
&&+(n-1) \biggl(-4 m \alpha(2s+1) - 3 d n (1 + n) \hbar^2 +  d
(4 \ell 
(\ell + 1)-4e^2g^2/\hbar^2 - 3 s (1 + s)) \hbar^2\\
&& + n (8 m \alpha+  6 d (1 + s) \hbar^2)\biggr)
a_{n-1} \\ 
&&+(n - s) \left((n-s)^2- (2\ell+1)^2+4e^2g^2/\hbar^2\right) \hbar^2
a_n\,,
\end{eqnarray*}
in which the coefficient $n-1$ does not cancel out as before (for $s=2\ell+2$)
because the last coefficient no longer factorizes in the same way.  In the
previous section we have already indicated several steps in the derivation of
the standard hydrogen spectrum that would no longer hold if $\ell$ (or the
effective $\tilde{\ell}$ in (\ref{tildel}) if $g\not=0$) is not an integer.

More specifically, we again now look at the case of $\ell=0$ or $s=2$,
comparing with the discussion at the end of the preceding section. Now,
choosing $n=1$ implies a non-trivial condition, given by $a_1=0$, because we
are no longer able to factor our $n-1$. With this value, $n=2$ is then
identically satisfied. At this stage, we have the same behavior as before,
with a single coefficient ($a_1$ here corresponding to $c_{-1}$ before)
required to be zero. At $n=3$, we obtain a linear relationship between $a_2$
and $a_3$, specifically
\begin{equation}
 2(m\alpha-de^2g^2) a_2=e^2g^2a_3\,.
\end{equation}
The previous equation, $c_0=0$, would correspond to $a_2=0$, which is implied
only if $g=0$, while $a_3=0$ may be implied for suitable quantized charges
such that $e^2g^2$ is an integer, given the value of $d$. For generic magnetic
charges $g$, and in particular for small ones such that
$0\not=e^2g^2/\hbar^2\ll 1$, $a_2$ and $a_3$ are not independent. It is then
impossible to make the recurrence end with a non-zero expectation value of
$e^{-k\hat{r}}$, which is a contradiction. As in the main text, we see that
the quantum number $\ell=0$ is ruled out for weak magnetic charges.

\end{appendix}

%\bibliographystyle{../preprint}
%\bibliography{../Bib/QuantGra}

\end{document}